\documentclass[twocolumn,english,pra,superscriptaddress]{revtex4-1}
\usepackage[utf8]{inputenc}
\usepackage{enumitem}
\usepackage{amsfonts}
\usepackage{amssymb}
\usepackage{color}
\usepackage{graphicx}
\usepackage{textcomp}
\usepackage{cancel}
\usepackage{dcolumn}
\usepackage{bm}
\usepackage{multirow}
\usepackage{float}
\usepackage{mathrsfs} 
\usepackage{braket}
\usepackage{amsmath}
\usepackage{subfigure}
\usepackage{epstopdf}
\usepackage{lipsum} 
\usepackage[usenames,divipsnames,svgnames,table]{xcolor}
\usepackage[unicode=true,pdfusetitle,
bookmarks=true,bookmarksnumbered=false,bookmarksopen=false,
breaklinks=true,pdfborder={0 0 0},backref=false,colorlinks=true]{hyperref}
\hypersetup{linkcolor=NavyBlue,urlcolor=NavyBlue,citecolor=NavyBlue}
\usepackage{breakurl}

\begin{document}
\title {Verification of colorable hypergraph states with stabilizer test}

\author{Hong Tao}
\affiliation{MOE Key Laboratory of Fundamental Physical Quantities Measurement, PGMF and School of Physics, Huazhong University of Science and Technology, Wuhan 430074, China}

\author{Xiaoqian Zhang}
\affiliation{State Key Laboratory of Optoelectronic Materials and Technologies and School of Physics, Sun Yat-sen University, Guangzhou 510006, China}

\author{Lei Shao}
\affiliation{Zhejiang Institute of Modern Physics, Department of Physics, Zhejiang University, Hangzhou 310027, China}

\author{Xiaoqing Tan}
\email{ttanxq@jnu.edu.cn}
\affiliation{College of Information Science and Technology, Jinan University, Guangzhou 510632, China}

\begin{abstract}
Many-body quantum states, as a matter of fact, are extremely essential to solve certain mathematical problems or simulate quantum systems in measurement-based quantum computation. However, how to verify large-scale quantum states, such as hypergraph states, is an exceedingly hard task for multi-body quantum systems. Here, we propose a novel fault-tolerant solution for the verification of colorable hypergraph states by using the stabilizer test. Compared with the adaptive stabilizer test, our protocol is dramatically facilitating by making only Pauli-$X$ and Pauli-$Z$ measurements. As to appliance, it will be also applied to blind quantum computing.
\end{abstract}

\maketitle

\section{Introduction}

Quantum computing offers a new approach to solve the NP-hard problem. Compared with classical computing, quantum computing can achieve exponential acceleration and provide a reliable guarantee for the security of quantum information processing \cite{Nielsen,Hayashi,Raussendorf,Afaulttolerant,Threshold2007}. However, how to determine the security, correctness, and fault tolerance of computational tasks is a problem to be solved when using quantum computing to process quantum information. Therefore, a fault-tolerant and verifiable quantum computing is indispensable and will be the focus of scientific research. In recent years, due to the rapid development of quantum error correction techniques, there are an increasing number of schemes to effectively verify the correctness, security, and feasibility of quantum computation \cite{Tomoyuki2014Morimae,Tomoyuki2016,Morimae2015Quantum,Morimae2017Verification,Hayashi2015Verifiable,Verifiablefaulttolerance,Gheorghiu2017Verification,PosthocVerification,Takeuchi2018Verification,Resourceefficient,MasahitoHayashi,Li2019Efficient,Yu2019Optimal,Huangjun2019Adversarial,Optimalverification,Fujii2016Power,twoqubitpurestates}. These schemes are implemented based on graph states, such as verifying Hamiltonian basis states \cite{Morimae2015Quantum}, stabilizer tests \cite{Hayashi2015Verifiable}, adaptive stabilizer tests \cite{Takeuchi2018Verification}.

Graph states are an important quantum resource for quantum computing with topologically protected sequences \cite{Topological,Raussendorf2007Topological}. If the probability distribution of non-adaptive sequential single qubit on a graph state is sampled classically and efficiently, then the polynomial hierarchy collapses to level 3 or level 2 \cite{AverageCase,Classicalsimulation,Quantumcommutingcircuits}.
Hypergraph state \cite{Rossi,EffiHypergraphStates,Qu,Xiong}, as the generalization of the graph state, is equally efficient in the measurement-based quantum computation (MBQC)
\cite{Briegel2009,Raussendor2003,FinettiTheorem,Broadbent2010Universal,Broadbent2010Measurement,morimae2013secure,li2014triple,Morimae2013Blind,Tan2018quantum,briegel2009measurement}, such as Union Jack state, which is one of the universal resource states for MBQC. As a kind of important quantum resources, the correctness of the quantum states generated by quantum devices becomes particularly important. Therefore, the verification of quantum states is necessary, and how to verify the hypergraph states of complex structures has become a difficult problem for many researchers.

However, in real experiments, more types of noise will be encountered, including various relations between qubits~\cite{Wallman2015Estimating, Ball2016Effect,Dawson2006}, experimental equipment error, human estimation error, external environment (temperature), etc.
Since the traditional quantum state tomography scheme will not be able to achieve fault tolerance, we need to find a new scheme to verify quantum states and realize quantum computing without assuming the underlying quantum state noise. In other words,
Given a desired quantum state, whose quantum device generates a quantum state  \cite{MasahitoHayashi,Li2019Efficient,Yu2019Optimal} that produces states $\rho_{1},\rho_{2},\cdots,\rho_{n-1},\rho_{n}$ in the $n$ rounds, which can be written as $\rho_i$,
and $\rho_{i}=|\Psi\rangle\langle\Psi|$ for pure states where all $i \in \{ 1,2, \cdots, n \}$. At the same time, in the case of extra noise, the density matrix of its quantum states satisfies
 $\langle\Psi|\rho_{i}|\Psi\rangle\leq 1-\epsilon$, where $\epsilon$ is a very small number.
However, there is a fault-tolerant technique that can reduce errors caused by additional noise (including, experimental equipment errors, estimation errors, etc.). In this paper, we propose a feasible verification hypergraph scheme, which increases the feasibility of our verification scheme by increasing the fault tolerance and reducing the error generated in the experimental process.

The need for a verification protocol satisfies two properties, completeness (if the prover is honest, equivalently, the probability that the state in the quantum register is the ground state of the Hamiltonian and the verifier accepts the prover is greater than $1-\mathrm{exp}(-n)$) and soundness (if the probability distribution of the target state is close to 1 when the verifier accepts the prover). These two properties are mathematically equivalent to error-tolerant detectability. For wrong output, we can detect it with high probability. But this is without satisfying the fault tolerance, and must also be coupled with acceptability, which means we need accepting the computational results of the test with high probability in a realistic noisy environment.

Here we propose a verifiable colorable hypergraph state protocol satisfying completeness and soundness (no underlying quantum graph noise are assumed).
We define the set of correctable errors on a given resource state, any small amount of noise on the hypergraph causes rejection, the set of correctable errors defines a bound, and we accept the test results within the specified range. We use stabilizer test scheme to verify a given hypergraph state $|G\rangle$ by decomposing hypergraph states into graph states. If the scheme is extended to the noisy case, we can determine whether a given hypergraph state is a fault-tolerant resource state within the correctable set. In this case,  the test passes, it is necessary to ensure that the accuracy is sufficiently high (the output of the hypergraph copy of the device gets 99\% fidelity with 90\% probability). Our verification scheme is validity and the resource consumption is a polynomial of constant order $O(n^{\gamma(G)})$. Finally, our fault-tolerant verifiable scheme can also be applied to blind quantum computing.

This paper is organized as follows. In Sec.~\ref{Sec_hyer}, we introduced what is a hypergraph state. In Sec.~\ref{Sec_Cover},  we give the error set of colorable hypergraph states. In Sec.~\ref{Sec_Case}, we consider a generic scenario on a three-colorable hypergraph state and give a test scheme with stabilizer tests.
In Sec.~\ref{Sec_verification}, we propose a verification protocol for hypergraph states based on the stabilizer test. Finally, we give the discussion in the paper.

\section{Hypergraph state}\label{Sec_hyer}

\begin{figure}[tbp]
\centering
\includegraphics[width=8cm]{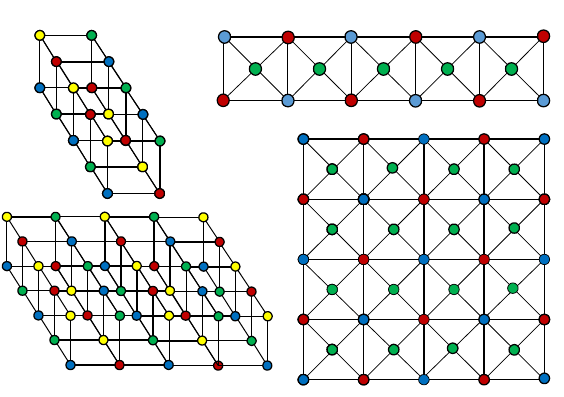}
\caption{ Examples of the hypergraph state: the Union Jack states on a 2-dimension lattice. The three vertices of each elementary are connected by an order-three hyperedge. The quantum states are three-colorable as illusrated.
\label{figframework}}
\end{figure}

We first define hypergraph state and describe their properties \cite{Morimae2017Verification}. The hypergraph $G=(V, E)$ is a pair of the set of vertices $V$ and the set of hyperedges $E$, where the numbers of qubits are $n = |V|$ and the edge satisfies $|e|\geq 2$ with $e\in E$, and $|e|$ is the number of edges linked to the hyperedge $e$. For example, the Union Jack state could be shown as Fig.~\ref{figframework}. The hypergraph state $|G\rangle$ corresponding to the hypergraph $G$ is defined by
\begin{eqnarray}
|G\rangle \equiv\left(\prod_{e \in E} \widetilde{\mathrm{CZ}}_{e}\right)|+\rangle^{\otimes n},
\end{eqnarray}
where $|+\rangle= \frac{1}{\sqrt{2}}(|0\rangle+|1\rangle)$, and $|0\rangle$ and $|1\rangle$ are the eigenstates  corresponding to the Pauli-$Z$ eigenvalues $\pm 1$, respectively. And the control-$Z$ operator $\widetilde{\mathrm{CZ}}_{e}$ can be written as
\begin{eqnarray}
  \widetilde{\mathrm{CZ}}_{e}\equiv \bigotimes_{i\in e}\openone_{i}-2\bigotimes_{i\in e}|1\rangle\langle1|_{i}
\end{eqnarray}
is the generalized $\mathrm{CZ}$ gate acting on vertices in the hyperedge $e$. Here, $\openone$ is the 2-dimension identity operator. For example, if $|e| = 2$, it is nothing but the standard CZ gate. If $|e| = 3$, it is the $\mathrm{CZ}_{2}$ gate,
\begin{equation}
\mathrm{CZ}_2\equiv\left(\openone^{\otimes 2}-|11\rangle\langle 11|\right) \otimes \openone+|11\rangle\langle 11| \otimes Z.
\end{equation}
 The density matrix form of the hypergraph state $|G\rangle$ is
\begin{equation}
\rho=|G\rangle \langle G|=\prod_{i=1}^{n} \frac{\openone^{\otimes n}+g_{i}}{2},
\end{equation}
 where the $i$th stabilizer $g_{i}$ of the hypergraph state $|G\rangle$ $(i =1,2,\cdots,n)$ is defined by
\begin{equation}
g_{i} \equiv\left(\prod_{e \in E} \widetilde{\mathrm{CZ}}_{e}\right) X_{i}\left(\prod_{e \in E} \widetilde{\mathrm{CZ}}_{e}\right).
\end{equation}
And the stabilizer $g_i$ satisfies the following property,
\begin{align}
  g_i |G\rangle = |G\rangle.
\end{align}

Hypergraph states have several advantages. For example, certain hypergraph states, such as the Union Jack state are universal resource states for measurement-based quantum computation with only Pauli measurements. With the development of quantum computers, it is of great significance to study the verification of hypergraph states. Some verification schemes for hypergraphs state have been proposed so far, and we present a relatively simple scheme to verify hypergraphs state.

\section{Cover strategy}\label{Sec_Cover}

A set $\mathcal{A}=\{\mathcal{A}_{1}, \mathcal{A}_{2}, \cdots,\mathcal{A}_{m}\}$ of the independent sets of $G$ is an  independent cover if $\bigcup\limits_{l=1}^{m} \mathcal{A}_{l}=V$. The cover $\mathcal{A}$ is also defined as the colorable set of the $G$ with $m$ colors when $\mathcal{A}$ is formed by the vertex $V$ (assuming $\mathcal{A}_{l}$ is a nonempty set of $V$). A hypergraph $G$ is $m$-colorable if its vertices can be colored using $m$ different colors. For
example, a 2-colorable graph is also called the bipartite graph. The chromatic number $\gamma(G)$ is the minimum number of colors in any coloring of the hypergraph $G$,
equivalently, the minimum number of elements in any independent cover of the hypergraph $G$, there is $\gamma(G)\leq m$, (when the hypergraph $G$ is geometric, for example, the Union Jack state, the chromatic number is equal to number of the colors $m$,  $\gamma(G)= m$.)

The $m$-colorable hypergraph state $|G\rangle$ is composed of the color-system $\mathcal{H}_{c_1}, \mathcal{H}_{c_2}, \cdots , \mathcal{H}_{c_m}$, consisting of $n_{c_1},n_{c_2}, \cdots, n_{c_m}$ qubits, where $n=n_{c_1}+n_{c_2}+ \cdots +n_{c_m}$, $n$ is the qubits number of hypergraph state $|G\rangle$, where $m \leq n$. For example, the Union Jack state can be shown in Fig.~\ref{Test}.
For each $\mathcal{A}_{l}$, each qubit on the hypergraph acts either as Pauli-$X$ operator or as Pauli-$Z$ operator. Depending on the color, it can be defined as the following two operator forms.  When $m$ is odd,
\begin{eqnarray*}\label{euX}
  X: = \bigotimes\limits_{i=1}^{n_{oo}} X_i, ~~~
  Z: = \bigotimes\limits_{i=1}^{n_{oe}} Z_i,\\
  X: = \bigotimes\limits_{i=1}^{n_{oe}} X_i, ~~~
  Z: = \bigotimes\limits_{i=1}^{n_{oo}} Z_i,
\end{eqnarray*}
where $n=n_{oo}+n_{oe}$, and $n_{oo} = n_{c_1}+n_{c_3}+\cdots+n_
{c_m}, ~ n_{oe}=n_{c_2}+n_{c_4}+\cdots+n_{c_{m-1}}$. Here, $n_{oo}, n_{oe}$ refer to the qubits number of subgraph in the subspace $\mathcal{H}_{c_1} \otimes \mathcal{H}_{c_3}, \cdots, \otimes \mathcal{H}_{c_m}$ and $\mathcal{H}_{c_2} \otimes \mathcal{H}_{c_4}, \cdots ,\otimes \mathcal{H}_{c_{m-1}}$, respectively.
And the operator $X_i$ $(Z_i)$ is Pauli-$X$ ($Z$) operator of the  $i$th qubits on the subgraph of the color-system $\mathcal{H}_{c}$ on the hypergraph state.
Similarly, when $m$ is even,
\begin{eqnarray*}\label{equX}
  X: = \bigotimes\limits_{i=1}^{n_{eo}} X_i,~~~
  Z: = \bigotimes\limits_{i=1}^{n_{ee}} Z_i,\\
  X: = \bigotimes\limits_{i=1}^{n_{ee}} X_i,~~~
  Z: = \bigotimes\limits_{i=1}^{n_{eo}} Z_i,
\end{eqnarray*}
where $n=n_{eo}+n_{ee}$, and $ n_{eo}=n_{c_1}+n_{c_3}+\cdots+n_{c_{m-1}}, ~ n_{ee}=n_{c_2}+n_{c_4}+\cdots+n_{c_m} $.

However, there are two tests for $\mathcal{A}_{l}$ on system $\mathcal{H}_{c_1}\otimes \mathcal{H}_{c_2}\otimes \cdots \otimes\mathcal{H}_{c_m}$. We can divide hypergraphs $G$ into $\gamma(G)$ graphs. There are two kinds of operators $X$, $Z$ on the system $\mathcal{A}_{l}$. According to the stabilizer test scheme~\cite{Hayashi2015Verifiable},
there are a total of $\gamma(G)$ efficient graphs, the test number of stabilizer test is $\frac{\gamma(G)\times(\gamma(G)-1)}{2}$.
Now we consider the first open set $\mathcal{A}_{1}$.
The hypergraph state $|G\rangle$ is defined in the following two forms. When $m$ is odd, one can get the form
\begin{eqnarray}
  (X_{n_{c_1}}\otimes Z_{n_{c_2}}\otimes X_{n_{c_3}}\otimes \cdots \otimes X_{n_{c_m}})|G\rangle=|G\rangle,
\end{eqnarray}
\begin{eqnarray}
(Z_{n_{c_1}}\otimes X_{n_{c_2}}\otimes Z_{n_{c_3}}\otimes \cdots \otimes Z_{n_{c_m}})|G\rangle = |G\rangle,
\end{eqnarray}
where $X_{n_{c_j}}= \bigotimes\limits_{i=1} ^{n_{c_j}}X_i, ~ Z_{n_{c_j}}= \bigotimes\limits_{i=1} ^{n_{c_j}}Z_i$ show that perform the Pauli-$X$($Z$) operator on the colorable-qubit $c_j$.
Therefore, when $m$ is even,
\begin{eqnarray}
  (X_{n_{c_1}}\otimes Z_{n_{c_2}}\otimes X_{n_{c_3}}\otimes \cdots \otimes Z_{n_{c_m}})|G\rangle=|G\rangle,
\end{eqnarray}
\begin{eqnarray}
  (Z_{n_{c_1}}\otimes X_{n_{c_2}}\otimes Z_{n_{c_3}}\otimes \cdots \otimes X_{n_{c_m}})|G\rangle=|G\rangle.
\end{eqnarray}

We perform the fault-tolerant MBQC on the $m$-colorable hypergraph states. The total space $\mathcal{H}_{c_1}\otimes \mathcal{H}_{c_2}\otimes \cdots \otimes\mathcal{H}_{c_m}$
is spanned by $\{Z^{x}|G\rangle\}_{x\in \mathcal{F}^{n}_{2}}$ in Ref.~\cite{Verifiablefaulttolerance}.
However, the sets of the correction errors on the $m$-colorable hypergraph state are defined as that the correctable state $|G\rangle$ and the erroneous $Z^{x}|G\rangle$ cause the same the result under the error correction. The set of correction error is specific on the $\mathcal{F}^{n_{c_1}}_{2}\otimes \mathcal{F}^{n_{c_2}}_{2}\otimes\cdots\otimes \mathcal{F}^{n_{c_m}}_{2}$,
the subset of the set $S$ can be written as $S_{c_1}\otimes S_{c_2}\otimes \cdots \otimes S_{c_m}$.

\section{Case study}\label{Sec_Case}
\begin{figure}[tbp]
\centering
\includegraphics[width=8cm]{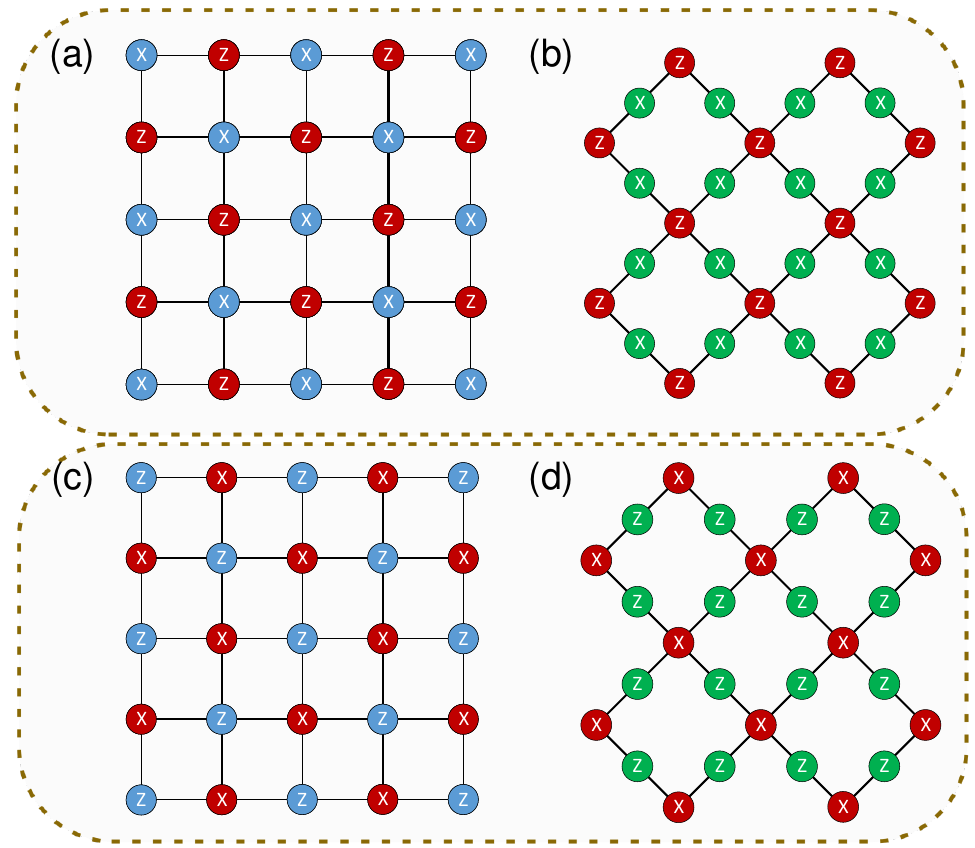}\\
\caption{The Union Jack states $G$, the decomposition structure of the total hypergraph states. The first step is applied the method of the stabilizer test to the first diagram (a, c), and the second step is also applied to the second diagram (b, d).
\label{Test}}
\end{figure}

Let us consider a generic scenario on a three-colorable hypergraph state $|\Psi_{BRG}\rangle$ composed of the blue system $\mathcal{H}_{B}$ , the red system $\mathcal{H}_{R}$, and the green system $\mathcal{H}_{G}$, consisting of $n_{B} , n_{R}, n_{G}$ qubits, where $n=n_{B}+n_{R}+n_{G}$. For example, the decomposition structure of the Union Jack state is shown as Fig.~\ref{Test}.
We can obtain the follow equations
\begin{eqnarray}\label{equ1}
  (X_{n_{B}}\otimes Z_{n_{R}}\otimes X_{n_{G}})|\Psi_{BRG}\rangle=|\Psi_{BRG}\rangle,
\end{eqnarray}
\begin{eqnarray}
  (Z_{n_{B}}\otimes X_{n_{R}}\otimes Z_{n_{G}})|\Psi_{BRG}\rangle=|\Psi_{BRG}\rangle,
\end{eqnarray}
where
\begin{align*}
  X_{n_{B}} = \bigotimes \limits_{i=1}^{n_B} X_i, ~~ Z_{n_{B}} = \bigotimes \limits_{i=1}^{n_B} Z_i, \\
  X_{n_{R}} = \bigotimes \limits_{i=1}^{n_R} X_i, ~~ Z_{n_{R}} = \bigotimes \limits_{i=1}^{n_R} Z_i, \\
  X_{n_{G}} = \bigotimes \limits_{i=1}^{n_G} X_i, ~~ Z_{n_{G}} = \bigotimes \limits_{i=1}^{n_G} Z_i
\end{align*}
are the Pauli operators on the blue (red, green) system.
We define the set of correction error $S$, the subset of $S$ is written as
\begin{eqnarray}\label{Eq:set}
  S_{B}\otimes S_{RG}, ~S_{R}\otimes S_{BG}, ~S_{G}\otimes S_{BR}.
\end{eqnarray}
By using the binary-valued adjacency matrix $A$, {i.e., $(i,j)$ element of $A$ is 1, if and only if the vertices $i$ and $j$ are connected} on the graph. For the Eq.~(\ref{Eq:set}), there are a number of relationships. For example, owning to $S_{B}\otimes S_{RG}$, we can get the following related equation
\begin{eqnarray}\label{equ2}
  X_{n_{B}}+ A_{1}^{T}Z_{n_{R}}+ A_{2}X_{n_{G}} \in S_{BG},
\end{eqnarray}
\begin{eqnarray}
  X_{n_{R}}+ A_{1}Z_{n_{B}}+ A_{2}^{T}Z_{n_{G}} \in S_{R},
\end{eqnarray}
where $A_1, A_2 $ are the adjacency matrices as shown in the Fig.~\ref{Test} (a) and (b). Then a set of correctable errors on the three-colorable hypergraph state $|\Psi_{BRG}\rangle$ is defined such that the correct state $|\Psi_{BRG}\rangle$ and an erroneous one $Z^x |\Psi_{BRG}\rangle$ result in the same computational outcome under the error correction.

\subsection{Test for verification quantum computation}

The text is similar to the stabilizer test procedure \cite{Hayashi2015Verifiable}. Using statistical sampling protocol to verify whether the error is correctable, and satisfies independent and identically distributed. The protocol process runs as follows

(1) The honest prover generated $|\Psi_{BRG}\rangle^{\otimes 6k+1}$, the prover sends each qubit to the verifier one by one.

(2) The verifier divides $6k+1$ blocks of $n$ qubits into four groups, $3\times 2k$ blocks and a single block (the fourth group), by random choice.

(3) The verifier uses the fourth group for her computation, other blocks are used for the test, which will be explained later.

(4) If the verifier passes the test, she accepts the result of the computation performed on the fourth group.

For each block of the first, the second, and the third groups, the verifier performs the following test.

Test for the blue system $T_{B}$ : For each two blocks of the first group, the verifier measures the qubits of the blue systems into the $X(Z)$ basis, respectively. Then she obtains the $
  X_{n_{B}}, ~Z_{n_{R}}, ~X_{n_{G}}$ and
  $Z_{n_{B}}, ~X_{n_{R}}, ~Z_{n_{G}}$.
If the following relationship is satisfied
\begin{eqnarray} \label{equ3}
  X_{n_{B}}+ A_{1}^{T}Z_{n_{R}}+ A_{2}X_{n_{G}} \in S_{BG},
  \end{eqnarray}
  \begin{eqnarray}
   A_{1}Z_{n_{B}}+X_{n_{R}}+ A_{2}^{T}Z_{n_{G}} \in S_{R},
\end{eqnarray}
then the test is passed.

Test for the red system $T_{R}$ : For each two blocks of the second group, the verifier measures the qubits of the red systems into the $X(Z)$ basis, respectively. Then she obtains the
$X_{n_{R}}, ~Z_{n_{G}}, ~X_{n_{B}}$ and
  $Z_{n_{R}}, ~X_{n_{G}}, ~Z_{n_{B}}$.
If the following relationship is satisfied
\begin{eqnarray}\label{equ4}
  X_{n_{R}}+ A_{1}^{T}Z_{n_{G}}+ A_{2}X_{n_{B}} \in S_{RB},
  \end{eqnarray}
  \begin{eqnarray}
  A_{1}Z_{n_{R}}+X_{n_{G}}+ A_{2}^{T}Z_{n_{B}} \in S_{G},
\end{eqnarray}
then the test is passed.

Test for the green system $T_{G}$ : For each two blocks of the third group, the verifier measures the qubits of the green systems into the $X(Z)$ basis, respectively. Then she obtains the
  $X_{n_{G}}, ~Z_{n_{B}}, ~X_{n_{R}}$ and
  $Z_{n_{G}}, ~X_{n_{B}}, ~Z_{n_{R}}$.
If the following relationship is satisfied
\begin{eqnarray}\label{equ5}
  X_{n_{G}}+ A_{1}^{T}Z_{n_{B}}+ A_{2}X_{n_{R}} \in S_{GR},
  \end{eqnarray}
  \begin{eqnarray}
  A_{1}Z_{n_{G}}+X_{n_{B}}+ A_{2}^{T}Z_{n_{R}} \in S_{B},
\end{eqnarray}
then the test is passed.

The probability $p_{\text {test }, i}$, that the verifier passes the stabilizer test for $g_i$ on the quantum state $\rho_{BR}$ is
\begin{eqnarray}
p_{\text {test}, i} =\frac{1}{2}+\frac{\operatorname{Tr}\left(\rho_{BR} g_{i}\right)}{2^{r+1}}.
\end{eqnarray}
Here we can see that if $r=\mathrm{poly}(n)$, then $p_{\text{test}, i}=\frac{1}{2}+O\left(2^{-\text {poly}(n) }\right)$\cite{Morimae2017Verification}. The probability
$P_{\text{pass}}$ is the expected probability that the verifier passes the test on the quantum state $\rho= |\Psi_{BRG}\rangle \langle\Psi_{BRG}|$, where the expectation passes the $N$-random sampling test. For the probability of the first resultant state $\rho_{BR}$ is
\begin{eqnarray}\label{equ6}\nonumber
  P_{\text {pass1 }}&=&\langle\Psi_{BR} | \rho_{BR} |\Psi_{BR}\rangle\\ \nonumber
  &=&\mathrm{Tr}(\rho_{BR} \prod\limits_{i=1}^{n_{B}+n_{R}}(\frac{\openone^{\otimes n_{B}+n_{R}}+g_{iB}}{2})) \\
  &\geqslant& 1-\frac{1-\delta}{\delta N }
\end{eqnarray}
with significance level $\delta$, $\delta \geq \frac{1}{N+1}$, and quantum state $|\Psi_{BR}\rangle \langle\Psi_{BR}|=\mathrm{Tr}_G(\rho)$.
Here $g_{iB}$ is the stabilizer on the graph state as shown in Fig.~\ref{Test} (a).
The probability of the second resultant state  $\rho_{RG}$ is
\begin{eqnarray}\label{equ7}
  P_{\text {pass2 }}&=&\langle\Psi_{RG} | \rho_{RG} |\Psi_{RG}\rangle \\
  &=&\mathrm{Tr}(\rho_{RG} \prod\limits_{i=1}^{n_{R}+n_{G}}(\frac{\openone^{\otimes n_{R}+n_{G}}+g_{iG}}{2})),\nonumber
\end{eqnarray}
where $g_{iG}$ is the stabilizer on the graph state as shown in Fig.~\ref{Test} (b), and quantum state $|\Psi_{RG}\rangle \langle\Psi_{RG}|=\mathrm{Tr}_B(\rho)$.
Due to the independence of the tests with $i= B, G$, the joint probability of passing the test can be written as
\begin{eqnarray}\label{equ9}
  P_{\text {pass}} > 1-\frac{2(1-\delta)}{\delta N }.
\end{eqnarray}
When the number of samples is large enough, we can say that our resulting state is our desired quantum state.
Next, now we introduce two important properties (detectability and acceptability)

\textsl{Detectability}, which is that if the error of the quantum device is not correctable, the fault output of the quantum computation is detected with high probability. According to Ref.~\cite{Hayashi2015Verifiable},
we can conclude the theorem in the same way. The verifier asks the prover to generate $6k+1$ copies $|\Psi_{BRG}\rangle^{\otimes 6k+1}$ of the hypergraph state $|\Psi_{BRG}\rangle$, where quantum state $|\Psi_{BRG}\rangle$ is an $n$-qubit hypergraph state and $k$ is the size of the polynomial $k=\mathrm{ploy}(n)$.

\textbf{Theorem 1 Detectability} If the test passes, with the significance level $\alpha$, assuming that $\alpha\geq \frac{1}{6k+1}$,  we can ensure the computable state $\rho_{com}$ of the fourth group satisfies
\begin{equation}
\mathrm{Tr}(\rho_{com} \mathrm{M}_{s})\geq 1-\frac{1}{\alpha(6k+1)},
\end{equation}
where $\mathrm{M}_{s}$ is the projector of the subspace.

According to the relation between Theorem 1 and the fidelity $\langle \Psi_{BRG}| \rho_{com} |\Psi_{BRG}\rangle$, and if the test passes, it is necessary to ensure that the accuracy is high enough, and the outputs copies of the hypergraph of the device have 90\% probability to get the target with 99\% the fidelity.
\begin{eqnarray}\label{equ12}
\langle \Psi_{BRG}|\rho_{com} |\Psi_{BRG}\rangle \geq \frac{1}{\sqrt{\alpha(6k+1)}}.
\end{eqnarray}
However, the property of the fault-tolerant verification protocol indicates that the probability of the computable result and the actual result is less than $\frac{1}{\sqrt{\alpha(6k+1)}}$. In other words, when $k$ is large enough, the verification is satisfied. Note that, if the prover would generate $6k$ correct copy state $\left( \left| \Psi_{BRG} \right\rangle \left\langle \Psi_{BRG} \right|\right)^{\otimes6k}$ of the hypergraph state and a single random wrong copy state $\rho_{\text{wrong}}$, then the probability that the prover will fool the verifier is $\frac{1}{6k+1}$. The above theorem does not assume the underlying graph state with the noise model. In the actual measurement, the noise will be taken on the hypergraph state. Even if the measurement results do not depend on the measurement basis, each qubit of the hypergraph state can be randomly rotated during the measurement. In such a case, the verification protocol will no longer be effective. Next, we should consider the acceptability, which ensures that the protocol will be applicative even if in the real measurement.

\textsl{Acceptability}. To calculate acceptability, that is, the given error subsets distribution $P$ is at the set
  $\mathcal{F}_{2}^{B}\otimes \mathcal{F}_{2}^{R}\otimes \mathcal{F}_{2}^{G}$
 of the $X$-basis or $Z$-basis errors. Then we denote the marginal distribution with the pair of $X$-basis on the blue system, $Z$-basis on the ren system, and $X$-basis on the green system ($Z$-basis on the blue system, $X$-basis on the red system, and $Z$-basis on the green system ). Here, the probability which the verifier passes the test $T_{B}, T_{R}, T_{G}$ with one round is
\begin{eqnarray*}
  P_{BG}(S_{BG})P_{R}(S_{R}), \\
  P_{RB}(S_{RB})P_{G}(S_{G}), \\
  P_{GR}(S_{GR})P_{B}(S_{B}).
\end{eqnarray*}
Since we need to perform $2k$ rounds on them,  the probability can be passed is
  \begin{eqnarray*}
   P_{BG}^{k}(S_{BG})P_{R}^{k}(S_{R})
   P_{RB}^{k}(S_{RB})P_{G}^{k}(S_{G})
   P_{GR}^{k}(S_{GR})P_{B}^{k}(S_{B}).
 \end{eqnarray*}
 Hence, if the probabilities $P_{BG}(S_{BG})$ and $P_{R}(S_{R})$, and $P_{RB}(S_{RB})$ and $P_{G}(S_{G})$, and $P_{GR}(S_{GR})$ and $P_{B}(S_{B})$ are close to 1, the verifier should accept the correct computation result on the fourth group with high probability.

 Compared with the other verification scheme of hypergraph state \cite{Morimae2017Verification}, our verifiable scheme is feasible, and its computational complexity is equivalent to that of graph states with stabilizer test \cite{Hayashi2015Verifiable,Takeuchi2018Verification}.
 By increasing fault tolerance with the error-correctable error sets, reducing the extra errors caused by underlying quantum graph state noise, the external environment and experimental equipment, increasing the verifier's trust in the prover, and improving the correctness of the quantum state. We can define a set of error-correctable error sets on the hypergraph states to improve the correctness of quantum states. For example, the schemes in Refs. \cite{Ball2016Effect,Ball2016PRA} and Refs. \cite{Afaulttolerant,Threshold2007} could be viewed using the surface code with the concatenated Reed-Muller 15-qubit code~\cite{Raussendorf2007,Fowler2009} and the concate-nated Steane 7-qubit code \cite{Fujii2015,Steane1999}.

\section{ Verification protocol}\label{Sec_verification}

\begin{figure}[tbp]
\centering
\includegraphics[width=8cm]{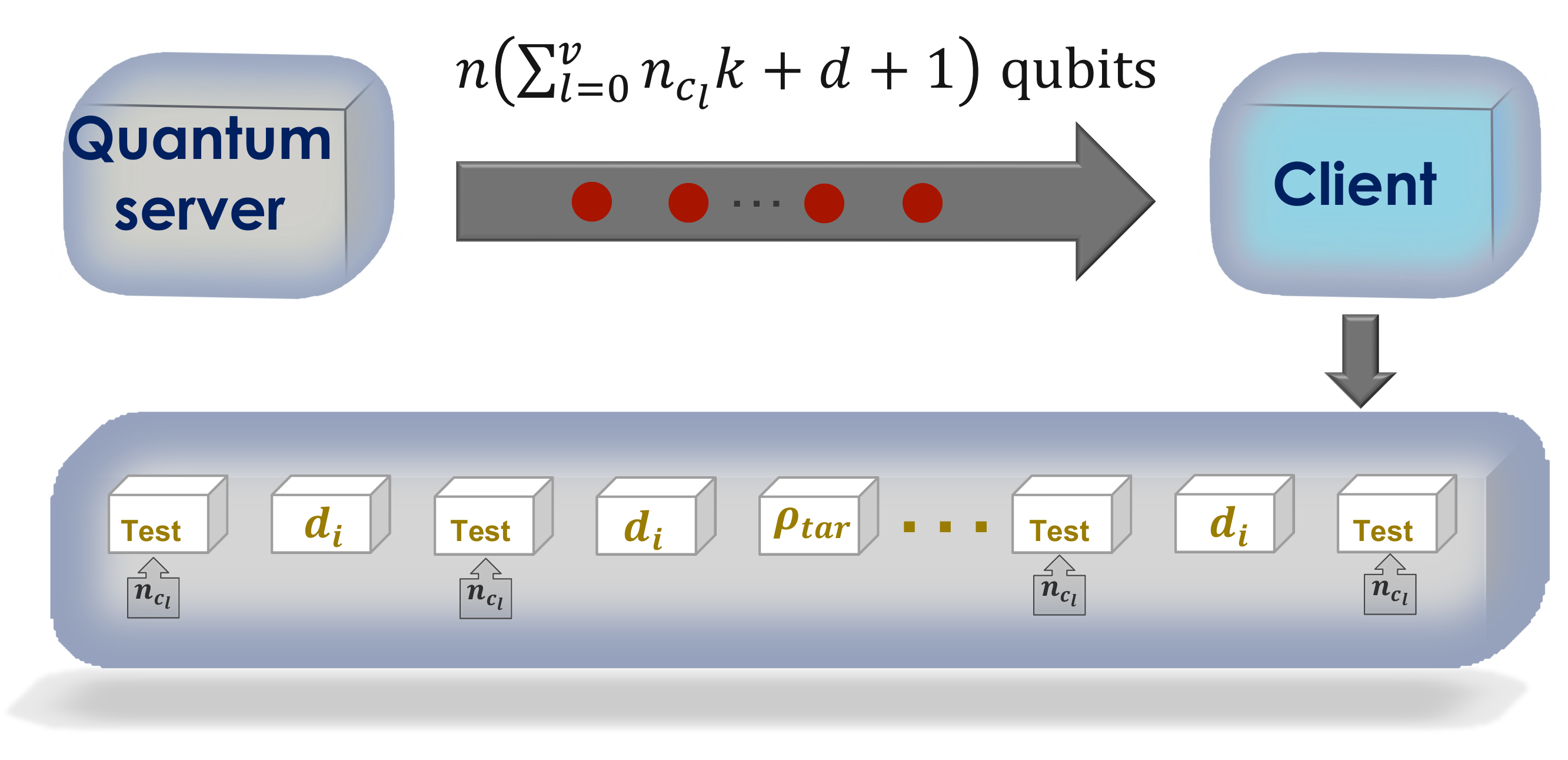}
\caption{ The verification scheme of the prover-verifier interaction considered in this paper. The quantum server is designated as the prover, and the client is designated as the verifier. The red dots represent qubits, and $n(\sum_{l=0}^{\upsilon}n_{c_l}k +d+1)$ is the number of qubits that the quantum server needs to transmit to the client.
\label{figVerification}}
\end{figure}

A weighted independence cover set $(\mathcal{A}, \mu)$ of the hypergraph state $\rho=|G\rangle \langle G|$ is a open cover with weights $\mu_{l}$ for $\mathcal{A}_{l}\in \mathcal{A}$ that is composed of $m$ nonempty independence sets, where $\mu_{l}$ is a probability distribution. We can structure a verification protocol for the $m$-colorable hypergraph state $\rho$. We propose a verification protocol for the hypergraph states based on the stabilizer test~\cite{Hayashi2015Verifiable}. In Fig.~\ref{figVerification}, we show the verification scheme. Our verifiable protocol runs as follows.

Step 1. The prover sends the verifier an $n(\sum_{l=0}^{\upsilon}n_{c_l}k +d+1)$-qubit quantum state, the quantum state consists of the $\sum_{l=0}^{\upsilon}n_{c_l}k +d+1$ registers, and each register stores $n$ qubits, where the $\upsilon$ is written as $\frac{\gamma(G)\times
(\gamma(G)-1)}{2}$, and $n_{c_l}= n_{\upsilon-l}$. If the prover is honest, the prover sends
$\rho^{ \otimes \sum_{l=0}^{\upsilon}n_{c_l}k +d+1}$ to the verifier.
On the other hand, if the prover is malicious, the prover sends an any $n(\sum_{l=0}^{\upsilon}n_{c_l}k +d+1)$ qubits instead of the correct quantum state $\rho^{\otimes\sum_{l=0}^{\upsilon}n_{c_l}k +d+1}$.

Step 2. The verifier chooses $d$ registers uniformly random and discards them to guarantee that the remaining $n(\sum_{l=0}^{\upsilon}n_{c_l}k +1)$-qubit state $\rho_{r}$ is close to an independent and identically distributed sample by using the quantum de Finetti theorem
 \cite{FinettiTheorem}.
Next, the verifier chooses one register called the target register with the target (correct) quantum state $\rho_{tar}$. The quantum state of the target register is uniformly random and used for the quantum computation. The remaining $\sum_{l=0}^{\upsilon}n_{c_l}k $ registers are divided into $\sum_{l=0}^{\upsilon}n_{c_l}$ groups such that which the $k$ registers are assigned to the every group is uniformly random. The verifier performs the parceled stabilizer test for $g_{ij}$ on the every register in the $i$th group, where $i \in \{ 1,2,\cdots, n_{c_i} \}, j\in \{1,2,\cdots, \upsilon\}$.

Let $K_{j}$ be the number of times that the verifier passes the parceled stabilizer test for $g_{ij}$ on $j$.  If every group of tests passes, then the following formula is satisfied
\begin{eqnarray}
\frac{K_{ij}}{K_{j}}\geq \frac{1}{2}+\frac{1-\epsilon}{2^{r+1}},
\end{eqnarray}
where $r=\mathrm{ploy}(n)$, and $n=n_{c_1}+n_{c_2}+n_{c_3}+...+n_{c_m}$,  $\epsilon=\frac{1}{n^{3}}, K_j=\frac{n^{7}r^{2}}{2}$.
We see that the verifier passes the stabilizer test for the $i$th group. If the verifier passes the stabilizer test for all $i$ and $j$, which means that the verifier accepts the prover.

The validity of the protocol is proved by showing the completeness and correctness of the protocol. Simply say, we show that the verification protocol is complete if the verifier accepts the correct quantum state with high probability. On the other hand, if the protocol can guarantee that the quantum state that passes the verification protocol has a high probability of being close to the rationally correct state, then the protocol is soundness.

\textbf{Theorem 2 (Completeness)} If the prover is honest, i.e., the quantum state of each register is the correct quantum state $\rho$, the probability that the verifier accepts the prover is $ \geq 1-\upsilon n e^{-\mathrm{ploy}(n)}$.

\textit{Proof.} The quantum state of each register is the correct quantum state $\rho$, the passing probability of the quantum state is $P_{\text {pass}}=\frac{1}{2}$, and the passing probability $p_{\text{test}, i}=\frac{1}{2}+O\left(2^{-(r+1) }\right)$ with stablizer test. Because of the Union bound and the Hoeffding inequality, we can obtain the probability which the verifier accepts the prover,
\begin{align} \label{equ16} \nonumber
  P_r
  &=P[\bigwedge\limits_{j=1}^{\upsilon}\bigwedge\limits_{i=1}^{n}(\frac{K_{ij}}{K_{j}}
  \geq \frac{1}{2}+\frac{1-\epsilon}{2^{r+1}})]\\ \nonumber
  &\geq 1-\sum\limits_{j=1}^{\upsilon}\sum\limits_{i=1}^{n}P[\frac{K_{ij}}{K_{j}}< P_{\text{test},i}-\frac{\epsilon}{2^{r+1}}]\\
  &\geq 1-\upsilon n e^{-\mathrm{ploy}(n)}.
\end{align}
The verifier accepts the prover is $ \geq 1-\upsilon n e^{-\mathrm{ploy}(n)}$. $\hfill\blacksquare$

\textbf{Theorem 3 (Soundness)} If the verifier accepts the prover, the state of the target register satisfies
\begin{eqnarray}\label{equ17}
\langle G|\rho_{tar}|G\rangle\geq 1-\frac{1}{n\upsilon},
\end{eqnarray}
with a probability larger than $1-\frac{1}{n\upsilon}$.

\textit{Proof.} Let $\mathrm{M} $ is the $n_{c_l}$-qubit projector $\mathrm{M}= \openone^{\otimes n_{c_l}}-|\mathcal{A}_{l}\rangle \langle \mathcal{A}_{l}|$,
Here $|\mathcal{A}_{l}\rangle$ represents subset of hypergraph states $|G\rangle$ with the weighted independence cover $(\mathcal{A}_l, \mu_l)$ for all $l$.
The set $\mathcal{A}$ of the hypergraph state $|G\rangle$ is an independence cover if $\bigcup\limits_{l=1}^{m}\mathcal{A}_{l}=V$.
$ T$ is the POVM element corresponding to the event where the verifier accepts the prover. We can show that any $n_{c_l}$-qubit state $\rho_{c_l}$ is in the independence cover $\mathcal{A}_{l}$. According to the results in Refs.~\cite{Morimae2017Verification,Takeuchi2018Verification},
\begin{eqnarray}\label{equ18}
\mathrm{Tr}\left[\left( T\otimes \mathrm{M} \right)\rho_{c_l}^{\otimes (n_{c_l}\cdot k+1)}\right]\leq\frac{1}{2n_{c_l}^{2}}.
\end{eqnarray}
Because of the quantum De Finetti theorem (for the fully one-way LOCC norm) \cite{FinettiTheorem} and Eq.~(\ref{equ18}), we could obtain
\begin{align} \label{equ19}  \nonumber
  \mathrm{Tr}\left[\left( T\otimes \mathrm{M}  \right)\rho_{r}\right]
  \leq &\mathrm{Tr}\left[\left( T\otimes \mathrm{M} \right)\int d\mu \rho^{\otimes \sum_{l=0}^{\upsilon}n_{c_l}k+1}\right]\\ \nonumber
  &+\frac{1}{2}
  \sqrt{
  \frac{2 \sum_{l=0}^{\upsilon} n^{3}_{c_l}  \upsilon^{3}k^{3} \mathrm{log}2}{d}} \\
  &\leq \frac{1}
  {(\sum_{l=0}^{\upsilon}n^{2}_{c_l})\upsilon^{2}},
\end{align}
where $d=2n^{7}_{c_l}\upsilon^{7}k^{2}\mathrm{log}2$.
Here, $\mu$ is a probability measurement on the quantum state $\rho$. There are the following facts
\begin{align}
  \sum_{l=1}^m n_{c_l}^2 \geq \frac{n^2}{m}=n^{2-\mathrm{log}_n m},
\end{align}
with $\gamma(G)\leq m, m\leq n$. Using Eq.~(\ref{equ18}) and Eq.~(\ref{equ19}) to compute that
\begin{align}\label{equ20}
  \mathrm{Tr}\left[\left( T\otimes \mathrm{M}  \right)\rho_{r}\right]
  &\leq\frac{1}{n^{2}\upsilon^{2}}.
\end{align}
Since we can have
\begin{equation}\label{equ21}
\mathrm{Tr}\left[( T\otimes \mathrm{M} )\rho_{r}\right]
=\mathrm{Tr}\left[( T\otimes \openone)\rho_{r}\right]\mathrm{Tr}\left[\mathrm{M} \rho_{tar}\right].
\end{equation}
Therefore, if
\begin{equation}\label{equ22}
\mathrm{Tr}[\mathrm{M} \rho_{tar}]> \frac{1}{n \upsilon},
\end{equation}
 then one can get
\begin{equation}\label{equ23}
\mathrm{Tr}[(T\otimes \openone)\rho_{r}]<\frac{1}{n \upsilon},
\end{equation}
which means that if the verifier accepts the prover,
\begin{equation}\label{equ24}
\langle G|\rho_{tar}|G \rangle\geq 1-\frac{1}{n \upsilon},
\end{equation}
with a probability larger than $1-\frac{1}{n \upsilon}$. $\hfill\blacksquare$

\section{application with Blind quantum computation}
In terms of application, we proposed the verification scheme to apply the blind quantum computation (BQC) \cite{Broadbent2010Universal,Broadbent2010Measurement,morimae2013secure,li2014triple,Morimae2013Blind,Tan2018quantum}. BQC allows a client with limited quantum technology
to delegate her quantum computational tasks to a server who can perform universal quantum computation while retaining the client’s secret information. The verifier (Alice) does not have enough quantum technology, and can delegate her quantum computing work to the prover (Bob), who has a full-fledged quantum computer (or perform the universal quantum computation), without leaking any her privacy.

Here, the quantum server generates $m$-colorable hypergraph states $\rho$ and sends them to the client with measurement-based quantum computing (MBQC)~\cite{briegel2009measurement} of verification scheme \cite{Hayashi2015Verifiable}.
According to the detectable, the exactness of
the quantum output is guaranteed under the acceptance, and the blindness is guaranteed by the no-signaling principle.  With the acceptance, our proposed verification scheme can apply blind quantum computation even under the quantum noise or quantum server deviation as long as the quantum server is honest. For the correct calculation state $\rho_{tar}$, Alice sends each qubit of the correct calculation state $\rho_{tar}$ to Bob one by one, and Bob executes the correct quantum calculation according to Alice's request. Likewise, if Bob is malicious, Alice does not accept the quantum state sent by Bob. The protocol is canceled. Assuming that Alice has a few quantum capabilities to perform single-qubit Pauli-$X$($Z$) measurements, then Alice can perform quantum computations on the correct calculation state $\rho_{tar}$.

\section{ discussion }
A graph state is a general quantum resource with quantum computation, and a hypergraph is a kind of graph state. Various schemes have been proposed for the verification of the graph state. However, due to the hypergraph state's more complex structure, the hypergraph state is more difficult to verify.
However, some verification schemes for hypergraphs have been proposed \cite{Morimae2017Verification,Huangjun2019Adversarial,Optimalverification} which are more complicated in form. Based on the stabilizer verification scheme of graph states in Ref.~\cite{Hayashi2015Verifiable}, a feasible scheme for verifying hypergraph states is proposed, our verification protocol requires weaker quantum computing technology for the verifier, and the verifier only need the power to perform the single-qubit measurement. Compared with other verification hypergraph schemes \cite{Huangjun2019Adversarial,Optimalverification}, the schemes established a general framework for verifying pure quantum states (including bipartite pure states, hypergraph states, Dicke states, GHZ states, stabilizer states, weighted graph states etc.). Our scheme adopts the stabilizer test scheme and only needs a single-qubit measurement.

Due to the more complex graph structure, the complexity of quantum computing is higher. Compared with graph states, the computational complexity of hypergraph state verification is higher. To reduce the complexity of the hypergraph state, we propose a fault-tolerant verification hypergraph protocol. Here, we compare the approach presented above with previous works. Our scheme does not increase the computational complexity, and the implementation scheme is relatively simple. we give a test scheme, decompose the hypergraph state into graph states, and test the correctness of the resulting quantum state using the stabilizer test scheme \cite{Hayashi2015Verifiable}. We give Theorem 1, which proves that our test scheme is feasible. The test and verification scenarios only need to perform single-qubit Pauli-$X$($Z$) measurements. The disadvantage is that more quantum resources are required.

In this paper, our verifiable protocol can be applied to arbitrary fault-tolerant hypergraph states (graph state) $|G\rangle$ in the MBQC. The fault-tolerant scheme reduces errors caused by underlying quantum graph state noise, the external environment and experimental equipment, increases the verifier's trust in the prover, and improves the feasibility of the verification scheme.  If the prover is honest, then the verifier can accept the correct computed state with a high probability greater than or equal to $1- \frac{1}{n \upsilon}$.
For a concrete example, we define a set of correctable errors based on topologically protected MBQC on Union Jack state. We can calculate the acceptability of distributed probability under the actual noise model. The probability of the test scheme passing with the test state $\rho$ is $P_{\text {pass}} > 1-\frac{2(1-\delta)}{\delta N }$, so we can accept test results with high probability.

\begin{acknowledgments}
This work was supported by the Natural Science Foundation of Guangdong
Province of China under Grant No. 2019A1515011069,  the Major Program of Guangdong Basic and Applied Research
under Grant No. 2019B030302008 and the National Natural Science Foundation of China under Grant Nos.
62032009 and 62005321.
\end{acknowledgments}

\appendix

\end{document}